\documentclass{iopart}
\usepackage{iopams}
\usepackage{bm}
\usepackage{graphicx}
\usepackage{hyperref}
\hypersetup{colorlinks,linkcolor=blue,filecolor=green,urlcolor=blue,citecolor=blue}

\def\bk{{\bf k}}

\def\br{{\bf r}}
\def\wk{\omega_k}
\def\wkp{\omega_{k'}}
\providecommand{\boldgreekletter}[1]{\mbox{\boldmath$#1$}}

\begin{document}

\title{Dynamical Casimir-Polder potentials in non-adiabatic conditions}
\author{Riccardo Messina$^1$, Roberto Passante$^2$, Lucia Rizzuto$^2$, Salvatore Spagnolo$^2$ and Ruggero Vasile$^3$}
\address{$^1$ Laboratoire Charles Coulomb UMR 5221 CNRS-UM2, D\'{e}partement Physique Th\'{e}orique, Universit\'{e} Montpellier 2, F-34095, Montpellier Cedex 5, France}
\address{$^2$Dipartimento di Fisica e Chimica, Universit\`{a} degli Studi di Palermo and CNISM, Via Archirafi 36, I-90123 Palermo, Italy}
\address{$^3$ Instituto de Fisica Interdisciplinar y Sistemas Complejos (IFISC), Universitat de las Illes Balears, E-07010 Palma de Mallorca, Spain}

\ead{lucia.rizzuto@unipa.it}
\begin{abstract}
In this paper we review different aspects of the dynamical Casimir-Polder potential between a neutral atom and a perfectly conducting plate under nonequilibrium conditions. In order to calculate the time evolution of the atom-wall Casimir-Polder potential, we solve the Heisenberg equations describing the dynamics of the coupled system using an iterative technique. Different nonequilibrium initial states are considered, such as bare and partially dressed states. The partially dressed states considered are obtained by a sudden change of a physical parameter of the atom or of its position relative to the conducting plate. Experimental feasibility of detecting the considered dynamical effects is also discussed.
\end{abstract}

\section{\label{sec:1}Introduction}

One of the most striking consequences of the quantum description of the electromagnetic field is the existence of quantum fluctuations of the electric and magnetic fields around their mean values, even in the vacuum state \cite{Milonni,CPP}.
Many important phenomena predicted by quantum electrodynamics and experimentally measured with remarkable precision, such as Lamb shift or spontaneous emission, are a consequence of vacuum field fluctuations \cite{Milonni,CPP,CT}.
As investigated by Casimir and Polder in 1948 \cite{C48,CP48}, quantum electrodynamics predicts also the existence of long-range interactions, having no classical analogue, between polarizable atoms or molecules, between atoms and neutral macroscopic objects, or between neutral macroscopic objects in vacuum.
The physical origin of these phenomena can be ascribed to the change of vacuum fluctuations due to the presence of boundary conditions for the field operators \cite{Milonni,PT82}.
Recent remarkably precise experimental results have confirmed the existence of the Casimir and Casimir-Polder forces \cite{Lamoreaux05,BKMM09}. These forces are also relevant in the design and implementation of micro- and nano-technologies \cite{BKMM09,DMRR}.
In recent years, particular attention has been given to the consequences of a nonadiabatic change of boundary conditions, material dielectric constants, atomic transition frequencies \cite{Dod10}.
It has been shown that in this case real photons can be produced, for example if conducting plates are put in motion with non-uniform acceleration in vacuum (dynamical Casimir effect) \cite{DMRR,Dod10}. Similar effects occur when the material properties (such as dielectric constant or plasma frequency, for example) of the macroscopic objects are appropriately changed \cite{Dod10,Pad09}, or in superconducting circuits by modulating the inductance of a quantum interference device at high frequencies \cite{WJP11}, or in Bose-Einstein condensates \cite{JPB12}.

A related effect is the dynamical Casimir-Polder force between two atoms or between an atom and a conducting mirror, when the system evolves starting from a nonequilibrium quantum state, for example a bare or partially dressed state, or an atomic excited state \cite{PP03,RPP04,VP08,MPRSV08,MVP10}.

In this paper, we present and review some recent results and relative physical interpretations about the dynamical Casimir-Polder potential between a perfectly conducting plate and a neutral two-level atom placed near it, considering different nonequilibrium initial states. In Section \ref{sec:2} we present the model used to perform the analysis of the systems under consideration. In Section \ref{sec:3}, we discuss our results relative to the time-dependent Casimir-Polder potential arising during the time evolution of an initially bare two-level atom placed near a perfectly conducting wall. In Section \ref{sec:4} we investigate a similar interaction when the system starts its unitary evolution from a partially dressed state. In Section \ref{sec:5}, finally, we present our results for the time evolution of the Casimir-Polder potential between a perfectly conducting plate and a neutral atom, when the position of the atom is suddenly changed at time $t=0$. Section \ref{sec:6} is devoted to some conclusive remarks.

\section{\label{sec:2}The Hamiltonian model}

We consider the Casimir-Polder potential between a two-level atom with transition frequency $\omega_0 = ck_0$ and a neutral perfectly conducting infinite plate. In the multipolar coupling scheme and within the electric dipole approximation, the atom-field interaction is given by $H_I =-\boldgreekletter{\mu}\cdot \bf{E}(\bf{r})$, where  $\boldgreekletter{\mu}$ and $\bf{E}(\bf{r})$ are respectively the atomic dipole moment and the transverse displacement field \cite{CPP,CT}; $\br=(x,y,z)$ is the position of the atom and the mirror coincides with the plane $z=0$. Thus the Hamiltonian describing our system reads
\begin{equation}
\label{Ham}
H=H_0+H_I,
\end{equation}
with
\begin{eqnarray}
\label{FreeIntHam}
H_0&=&\hbar\omega_0S_z+\sum_{\bk j}\hbar\wk a^{\dag}_{\bk j}a_{\bk j}\, , \nonumber \\
H_I&=&-i\sqrt{\frac{2\pi\hbar c}{V}}\sum_{\bk j}\sqrt{k}\left(\boldgreekletter{\mu}\cdot {\bf f}_{\bk j}(\br )\right)\left( a_{\bk j}-a^{\dag}_{\bk j} \right)
\left( S_+ + S_- \right) \, .
\end{eqnarray}
$S_z $, $S_+ $ and $S_{-}$ are the pseudo-spin operators of the two-level atom, and
\begin{eqnarray}
\label{ModeFunc}
({\bf f}_{\bk j}(\br ))_x &=& \sqrt{8}({\bf \hat{e}}_{\bk j})_x \cos[k_x (x+L/2)]\sin[k_y (y+L/2)]\sin[k_z z] \nonumber \\
({\bf f}_{\bk j}(\br ))_y &=& \sqrt{8}({\bf \hat{e}}_{\bk j})_y \sin[k_x (x+L/2)]\cos[k_y (y+L/2)]\sin[k_z z] \nonumber \\
({\bf f}_{\bk j}(\br ))_z &=& \sqrt{8}({\bf \hat{e}}_{\bk j})_z \sin[k_x (x+L/2)]\sin[k_y (y+L/2)]\cos[k_z z]
\end{eqnarray}
are the field mode functions evaluated at the atomic position $\br$, which take into account the presence of the wall (${\bf{\hat{e}}}_{{\bf k}j}$ are polarization unit vectors). The allowed values of $\bf{k}$ have components $(k_x,k_y,k_z)=(l\pi/L,m\pi/L,n\pi/L)$ with $l,m,n=0,1,2...$.
The mode functions in (\ref{ModeFunc}) are those of a perfectly conducting cubical cavity of volume $V = L^3$ with walls $-L/2<x<L/2$, $-L/2<y<L/2$, $0<z<L$. At the end we will take the limit $L\rightarrow\infty$, in order to turn from the cavity to the wall at $z = 0$ \cite{PT82}.

In order to calculate the time evolution of the Casimir-Polder atom-wall potential, we solve the Heisenberg equations of the atomic and field operators, using an iterative technique \cite{VP08,MVP10}. This method is based on the assumption that all atomic and field operators involved in the dynamics of the system can be  written as a power series in the coupling constant. So, the zeroth-order terms correspond to the absence of interaction and then to the free evolution given by $H_0$. For example, the field annihilation operator takes the form
\begin{equation}
a_{\bk j}(t)=a^{(0)}_{\bk j}(t)+a^{(1)}_{\bk j}(t)+a^{(2)}_{\bk j}(t)+...,
\end{equation}
where the contribution $a_{\bk j}^{(i)}(t)$ is proportional to the $i$-th power of the electric charge.
The second-order energy shift is given by \cite{MVP10}
\begin{equation}
\label{energyshift}
\Delta E^{(2)}(t)=\frac{1}{2}\langle\psi|H^{(2)}_{I}(t)|\psi\rangle,
\end{equation}
where $|\psi\rangle$ is the (time-independent) state and
\begin{eqnarray}
\label{Hit}
&\ & H^{(2)}_{I}(t)=-\frac{2\pi i c}{V}\sum_{\bk j} k\left( \pmb{\mu}\cdot {\bf f}_{\bk j}(\br)\right)^2[S_+e^{i\omega_0 t}+H.c.] \nonumber \\
&\times& \left\{ S_+[e^{-i\omega_{k}t}F(\omega_0+\omega_{k},t)
- e^{i\wk t}F^{\ast}(\wk-\omega_0,t)]-H.c.\right\} \nonumber \\
&+&\frac{4\pi i c}{V}S_z\sum_{\bk \bk 'jj'}\sqrt{kk'}\left( \pmb{\mu}\cdot {\bf f}_{\bk j}(\br )\right) \left( \pmb{\mu}\cdot {\bf f}_{\bk 'j'}(\br )\right)
[a_{\bk j}e^{-i\wk t}-H.c.]\nonumber\\
&\times& \left\{ a_{\bk 'j'}[e^{i\omega_0 t}F^{\ast}(\omega_0+\wkp ,t)-e^{-i\omega_0 t}F^{\ast}(\wkp -\omega_0,t)]+H.c.\right\}
\end{eqnarray}
is the interaction Hamiltonian at the second order in the coupling constant (all operators in the right-hand side of (\ref{Hit}) are taken at time $t=0$), and we have introduced the function $F(\omega,t)=(e^{i\omega t}-1)/(i\omega)$.

The main advantage of using the Heisenberg picture is that the operator dynamics is independent from the specific initial state of the system, and thus we can calculate the time evolution of any expectation value choosing the state of the system at the very end of the calculation.

\section{\label{sec:3}The bare state}

In this Section we present our results relative to the time-dependent Casimir-Polder potential arising during the time evolution of a two-level atom  at a distance $d$ from a perfectly conducting wall, assuming that at $t=0$ the coupled system is in its bare ground state. The bare ground state $|0_{\bk j},\downarrow\rangle$, where the atom is in its ground state and the field in the vacuum state, is the lowest-energy eigenstate of the unperturbed Hamiltonian $H_0$. It is a very useful idealization because it can give important hints on the behavior of realistic systems; however, it cannot be experimentally realized because the atom-field interaction cannot be switched off. Hence the necessity to introduce realistic partial dressed states, as we shall discuss in the next two Sections.
Using $|\psi \rangle=|0_{\bk j},\downarrow\rangle$ and substituting (\ref{Hit}) into (\ref{energyshift}), we finally obtain

\begin{eqnarray}
\label{Energyshift0}
\Delta E^{(2)}(t)&=&-\frac{\mu^2}{12\pi d^3}\lim_{m \to 1}D^m \nonumber \\
&\times& \bigg[\int_0^{+\infty} \! dk\frac{\sin(2mkd)}{k+k_0}\left[ 1-\cos \left((k_0+k)ct\right) \right]\bigg] \, ,
\end{eqnarray}
where $D^{m}=2-2\partial/\partial m+\partial^{2}/\partial m^2$ is a differential operator and we have assumed an isotropic atom. Eq. (\ref{Energyshift0}) gives an analytical expression of the time-dependent atom-wall Casimir-Polder potential energy. A remarkable property of the associated dynamical force is that, contrary to the usual stationary electric atom-wall Casimir-Polder force which is attractive for any atom-wall distance, Eq. (\ref{Energyshift0})  yields a force that oscillates in time from attractive to repulsive \cite{VP08}. Obtaining repulsive Casimir force is an important issue, in particular in the applications to devices such as microelectromechanical systems, where it could allow a better control of the movements of the parts of the device \cite{DMRR}.

\section{\label{sec:4}The partially dressed state (I)}

We now consider the case of a partially dressed atom, in order to overcome the limitations of the bare-atom case discussed in the previous Section. We consider an atom placed at $\br =(0,0,d)$, initially having a transition frequency $\omega'_0 = ck'_0$, and prepared in its dressed ground state, given by the following Eq. (\ref{dressedstate1}). We then suppose to produce at $t = 0$ a sudden change of its transition frequency, from $\omega_0'$ to a new frequency $\omega_0$, for example by the Stark shift due to an external electric field rapidly switched on. We assume that this change is so rapid that the quantum state immediately after $t = 0$ remains the same as before. Here, we take as initial state
\begin{equation}
\label{dressedstate1}
|\psi(0)\rangle=|0_{\bk j},\downarrow\rangle-i\sqrt{\frac{2\pi}{\hbar V}}\sum_{\bk j}\frac{\sqrt{k}\left( \pmb{\mu}\cdot{\bf f}_{\bk j}(\br)\right)}{k+k'_0}|1_{\bk j},\uparrow\rangle,
\end{equation}
that is the fully dressed ground state of the atom-field coupled system, as obtained by second-order perturbation theory. However, this state is not an eigenstate of the new Hamiltonian for $t > 0$ (where the new atomic frequency $\omega_0$ appears), so it will evolve over time (dynamical self-dressing), subjected to the unitary evolution given by the Hamiltonian (\ref{FreeIntHam}).
This evolution yields a time-dependent Casimir-Polder potential between the atom and the wall. In fact, substituting (\ref{Hit}) and (\ref{dressedstate1}) in (\ref{energyshift}), and neglecting higher-order terms, we obtain \cite{MVP10}
\begin{eqnarray}
\label{Energyshift1}
&\ &\Delta E^{(2)}(t)=-\frac{\mu^2}{12\pi d^3}\lim_{m \to 1}D^m\bigg[\int_0^{+\infty} \! dk\frac{\sin(2mkd)}{k+k_0}\nonumber\\
&\ &\times \left[1-\cos\left( (k_0+k)ct \right) \right]+\int_0^{+\infty} \! dk\frac{\sin(2mkd)}{k+k'_0} \cos\left( (k_0+k)ct\right) \bigg] \, ,
\end{eqnarray}
where for simplicity we have assumed the spatial isotropy of the atomic dipole, and the continuum limit ($V=L^3 \rightarrow \infty$) has been taken.

In order to produce the assumed change of the atomic transition frequency, the Stark effect can be exploited by subjecting at $t = 0$ the atom to a uniform electric field.
We have assumed that the frequency change is so rapid that the quantum state immediately after $t = 0$ remains the same as before. This nonadiabatic hypothesis is realistic if the time required to switch on the electric field is small compared to $\omega_{0}^{-1}$, which is a typical timescale of the atomic evolution.
An excellent candidate for the elaboration of an experimental scheme aiming  to investigate the system analyzed here is a Rydberg atom. Rydberg atoms typically have a transition frequency of some GHz and, in this case, switching on an appropriate electric field in times shorter than $\tau \simeq 10^{-9}$ s should be experimentally achievable (see for example \cite{PDG}), and thus our nonadiabatic hypothesis should be valid.
The dynamical Casimir-Polder force that can be obtained from Eq. (\ref{Energyshift1}) as minus its derivative with respect to the atomic position $d$, also shows oscillations from attractive to repulsive, and asymptotically ($t \rightarrow \infty$) it settles to the same asymptotic value for the initial bare state case, coinciding with the value obtained in the stationary case.

\section{\label{sec:5}The partially dressed state (II)}

In this section we present our results on the time evolution of the Casimir-Polder potential energy between a perfectly conducting plate and a neutral atom, that at time $t=0$ is suddenly moved to a different position. This sudden displacement of the atomic position generates a partially dressed state, because the atom-mirror distance has changed. For generality, we also assume a change of the atomic transition frequency, similarly to the case discussed in the previous Section.  So, up to time $t=0$ the atom has position $\bf{r}'$, transition frequency $\omega'_0=ck'_0$, and it is in its dressed ground state. At $t = 0$ a sudden change of the atomic position from $\bf{r}'$ to $\bf{r}$ occurs, as well as a rapid change of its transition frequency from $\omega_0'$ to the new frequency $\omega_0$. We assume both changes are so rapid that the quantum state immediately after $t = 0$ is the same as before. The dressed initial state
\begin{equation}
\label{dressedstate2}
|\psi(0)\rangle= |0_{\bk j},\downarrow\rangle-i\sqrt{\frac{2\pi}{\hbar V}}\sum_{\bk j}\frac{\sqrt{k}\left( \pmb{\mu}\cdot {\bf f}_{\bk j}(\bf{r}')\right)}{k+k'_0}|1_{\bk j},\uparrow\rangle,
\end{equation}
is not an eigenstate of the new Hamiltonian valid for $t > 0$ (where the new atomic frequency $\omega_0$ and the new atomic position $\bf{r}$ appear), so it will evolve over time. Let us note that in the the state (\ref{dressedstate2}), the old atomic position appears. Explicit evaluation of the interaction energy (\ref{Hit}) in this case, yields \cite{PRS13}
\begin{eqnarray}
\label{Energyshift2}
&\ &\Delta E^{(2)}(t)=-\lim_{m \to 1}\bigg\{\frac{\mu^2}{12\pi d^3}D^{m}\int_0^{+\infty}dk\frac{\sin(2mkd)}{k+k_0}
\Biggl[1\nonumber\\
&- &\cos \left( (k_0+k)ct \right) \Biggr]
+\frac{\mu^2}{12\pi \bar{z}^3}D^{m}\int_0^{+\infty}dk\frac{\sin(2mk\bar{z})}{k+k'_0}
\cos \left( (k_0+k)ct \right)\nonumber\\
&+&\frac{\mu^2}{12\pi z^3}D^{m}\int_0^{+\infty}dk\frac{\sin(2mkz)}{k+k'_0}\cos \left( (k_0+k)ct \right)\bigg\} \,
\end{eqnarray}
where $d=\mid \br\mid$, $d'=\mid \br ' \mid$, $\bar{z}=(d+d')/2$ and $z=(d-d')/2$. It is straightforward to verify that, if $d=d'$, Eq. (\ref{Energyshift2})  reduces to (\ref{Energyshift1}), as expected. Numerical evaluation of (\ref{Energyshift2}) shows that also in this case the force oscillates in time from attractive to repulsive, with features similar to those of the previous cases.
Analogously to the case discussed in the previous Section, it is important to verify if our hypothesis that the quantum state of the system remains unchanged immediately after the change of atomic position is physically correct and experimentally feasible.
Our nonadiabatic hypothesis is reasonable if the time required to change the position of the atom is small compared to $\omega_{0}^{-1}$; for a Rydberg atom this time is of the order of $10^{-9}$ s. For example, the thermal speed of a Rubidium atom at $300$ K is $300$ m/s, \cite{RHTN}. So, in $10^{-9}$s it can cover a distance of $3\times10^{-7}$ m, compatible with the dimensions of the modern cavities used in laboratories. This indicates that the thermal motion of the atoms could give rise to the dynamical effects here considered.
\section{\label{sec:6}Conclusions}
In this paper we have presented and reviewed some specific aspects of the dynamical atom-wall Casimir-Polder potential under non-adiabatic conditions, using different initial states of the atom-field system. Our method uses an iterative solution of the Heisenberg equations of motion for atomic and field operators. This method has the remarkable advantage that the specific initial state of the system must be specified only at the end of the calculation, and thus it makes easier to consider different initial states.
We have first considered the case of an initially bare ground state atom and found the time evolution of the atom-wall Casimir-Polder interaction, showing oscillations from attractive to repulsive character. We have also discussed the limits of this model based on a bare state. Then we have considered two different cases of more realistic partially dressed states, obtained by a sudden change of the atomic transition frequency and/or its position with respect to the conducting mirror. Features of the dynamical Casimir-Polder interactions, as well as possibility of detecting experimentally these new dynamical effects, have been discussed.

\section{Acknowledgments}
The authors gratefully acknowledge financial support by the Julian Schwinger Foundation, by Ministero dell'Istruzione, dell'Universit\`{a} e della Ricerca and by Comitato Regionale di Ricerche Nucleari e di Struttura della Materia.

\end{document}